\documentclass{IR-Template/ISAS_IR}

\usepackage[cmex10]{amsmath}
\usepackage{amsfonts}
\usepackage[figure]{algorithm2e}
\usepackage{ISASPackages/ISASMatheMacros}
\usepackage{abkuerzungen}
\usepackage{graphicx}
\usepackage{epstopdf}
\usepackage{subcaption}
\usepackage{flushend}
\usepackage{tabularx}
\usepackage{microtype}

\usepackage{amsthm}
\newtheorem{definition}{Definition}

\begin{document}

\begin{frontmatter}

\title{Unscented Orientation Estimation Based on the Bingham Distribution}

\author[isas]{Igor~Gilitschenski}
\ead{gilitschenski@kit.edu}

\author[isas]{Gerhard~Kurz}
\ead{gerhard.kurz@kit.edu}

\author[ucl]{Simon~J.~Julier}
\ead{s.julier@ucl.ac.uk}

\author[isas]{Uwe~D.~Hanebeck}
\ead{uwe.hanebeck@ieee.org}

\address[isas]{Intelligent Sensor-Actuator-Systems Laboratory (ISAS)\\
Institute for Anthropomatics\\
Karlsruhe Institute of Technology (KIT), Germany\vspace{3mm}}

\address[ucl]{Virtual Environments and Computer Graphics Group\\
Department of Computer Science\\
University College London (UCL), United Kingdom}

\begin{abstract}
Orientation estimation for 3D objects is a common problem that is usually tackled with traditional nonlinear filtering techniques such as the extended Kalman filter (EKF) or the unscented Kalman filter (UKF). Most of these techniques assume Gaussian distributions to account for system noise and uncertain measurements. This distributional assumption does not consider the periodic nature of pose and orientation uncertainty. We propose a filter that considers the periodicity of the orientation estimation problem in its distributional assumption. This is achieved by making use of the Bingham distribution, which is defined on the hypersphere and thus inherently more suitable to periodic problems. Furthermore, handling of non-trivial system functions is done using deterministic sampling in an efficient way. A deterministic sampling scheme reminiscent of the UKF is proposed for the nonlinear manifold of orientations. It is the first deterministic sampling scheme that truly reflects the nonlinear manifold of the orientation.
\end{abstract}

\end{frontmatter}

\section{Introduction}\noindent
Estimating orientation is fundamental for mobile sensor systems. In applications that range from UAVs to augmented reality, accurate orientation is required for tracking and control. In prepared environments, high precision estimation results can be achieved by using high cost installations integrated in the infrastructure. In many interesting scenarios, such as disaster relief or outdoor tracking, orientation estimation techniques cannot rely on a prepared environment. Additionally, it is often necessary to avoid the use of high cost sensors and rely on sensors with poor performance. Thus, there is a need to handle orientation estimation in the presence of high uncertainties. Stochastic filtering techniques can be used to estimate the true system state from noisy measurements.  

In this work, we consider a filter which removes the assumption of Gaussian noise and makes use of directional statistics for a better description of uncertainty in periodic data. This promises better results because certain assumptions motivating classical filtering approaches do not hold in a periodic setting. First, the Gaussian distribution is defined in the Euclidean space and thus does not consider periodicity. Second, Gaussians are preserved under linear operations. Unfortunately, there is no equivalent to a linear function on the manifold of orientations. Third, the central limit theorem usually motivates the use of Gaussians but it does not apply to certain manifolds such as the hypersphere $S^n$. 
Consequently, Gaussians are not a good approximation for periodic data in case of strong noise \cite{Mardia1999}. On a circle, for instance, the true limit distribution might be the wrapped normal distribution (which differs in its shape from the classical Gaussian distribution). 

\subsection{Recursive Orientation Estimation}

There has been lots of work on orientation estimation, but almost all of the proposed methods assume that the distributions are Gaussian. They are usually based on the extended Kalman filter (EKF) or the unscented Kalman filter (UKF) \cite{Julier2004}. Particularly estimation involving a unit quaternion representation of the system state is discussed in a broad range of literature, e.g., \cite{Kraft2003} and \cite{LaViola2003}. Quaternions are popular as they do not suffer from singularities or the ``gimbal lock'' (which can be suffered by Euler angles when one orientation axis is lost). A lot of the current approaches use nonlinear projection to push the state estimate back onto the surface of the sphere of unit quaternions.

Rather than using Gaussians and applying projection operations, we propose to use forms of uncertainty that explicitly describe the structure of the nonlinear manifold. Specifically, we propose to develop recursive estimators for orientation using the Bingham distribution, which is an antipodally symmetric distribution defined on the hypersphere \cite{Bingham1974}. This approach is used to describe uncertainty on the group of rotations $SO(3)$ parametrized by quaternions. Antipodal symmetry of the Bingham distribution accounts for the fact that the unit quaternions $q$ and $-q$ represent the same rotation. The proposed approach takes the periodic nature of the problem into account by making use of the periodic nature of the Bingham distribution.  

The use of directional statistics in recursive filtering has lately been discussed in the context of angular estimation \cite{Azmani2009, ACC13_Kurz, Fusion13_Kurz-Bingham, Stienne2013}. In \cite{Glover2011}, the Bingham distribution was used for Monte Carlo pose estimation. In \cite{Fusion13_Kurz-Bingham}, we proposed a recursive Bingham filter. It considers system functions performing a predefined change of orientation. They can be thought of as an equivalent to an identity system function in $\R^n$ with additive (possibly non-zero mean) stochastic noise. This approach was also independently developed in \cite{Glover2013}, which additionally considers an  constant velocity model for orientations.

\subsection{Main Contribution}
In practical applications, the system function might have a strong impact on the dispersion of the prior. In order to take this into account, we extend existing work on Bingham filtering by proposing an equivalent of the unscented Kalman filter (UKF) for orientations based on the Bingham distribution. Thus, the system state is an orientation represented directly in terms of quaternions. The new filter is based on two consecutive steps. First, the prediction step is performed by deterministically sampling the current system state and propagating the samples through the system function. Afterwards, system noise is imposed and a corresponding Bingham distribution is found by moment matching.  
The measurement update step assumes noisy measurements of the system state. It makes use of the fact that the product of two Bingham density functions is itself a rescaled Bingham density. Thus, the measurement update step can be performed in closed form. One of the challenges involved in handling the Bingham distribution is the computation of its normalization constant, which is discussed in \cite{Kume05, Koyama2012, Sei2013, Koev2005}. We will use an approach based on precomputed lookup-tables from \cite{Glover13}.

The remainder of the paper is structured as follows. In Sec.~\ref{sec:bingham}, we review the Bingham distribution and then propose a method for deterministic sampling in Sec.~\ref{sec:sampling}. The presented approach can easily be generalized to higher dimensions. Our filter is presented in Sec.~\ref{sec:filter}, where we make use of moment matching for quaternion multiplication in the prediction step and present a closed-form measurement update step. The proposed filter is compared against the UKF and the particle filter in Sec.~\ref{sec:evaluation}. Our work is concluded in Sec.~\ref{sec:conclusion}.

\section{Bingham Distribution}\label{sec:bingham}\noindent
The Bingham distribution naturally arises when conditioning a $d$-dimensional Gaussian distribution to the $d$-dimensional unit sphere (which is denoted by $S^{d-1}$). That is, a Gaussian random vector follows a Bingham distribution given its length is known to be 1. This distribution is of particular interest due to the following two facts. First, it offers a very natural way to describe uncertainty over unit quaternions and can be easily used for describing uncertain orientations. Second, it is closed under Bayesian inference, making efficient algorithms for recursive filtering possible. 

\begin{definition}
The probability distribution with probability density function
\begin{align*}
f(\ux; \fM, \fZ) = \fr{1}{N(\fZ)} \exp(\ux^T \fM\fZ\fM^T\ux)\ ,
\end{align*}
where $\fZ$ is a diagonal matrix with increasing entries $z_1\leq z_2 \leq .. \leq z_d$, $\fM$ is an orthogonal matrix, and $N(\fZ)$ is a normalization constant, is called the Bingham distribution. If $\uy$ is a Bingham distributed random vector with parameters $\fM$ and $\fZ$, we write $\uy \sim\binghamd(\fM,\fZ)$.
\end{definition}

This definition yields the normalization constant, which is given by
\begin{align*}
N(\fZ) = 
\int_{S^{d-1}} \exp\li(\ux^T \fZ \ux \ri) \df \ux\ .
\end{align*}
The matrix $\fM$ is not a part of the normalization constant, because $N(\fZ)=N(\fM \fZ\fM^T)$. It is also possible to represent the normalization constant in terms of a hypergeometric function of matrix argument $ _1F_1(\cdot, \cdot, \cdot)$ \cite{Herz1955}, that is
\begin{align*}
N(\fZ) = |S^{d-1}|\cdot\ _1F_1\li(\fr{1}{2},\fr{d}{2}, \fZ \ri)\ ,
\end{align*}
where $|S^{d-1}|$ is the surface area of the unit ball in $\R^d$. Computation of the normalization constant is difficult and gets harder as the dispersion decreases. Approaches to solve this problem include series expansions \cite{koev2006}, saddle point approximations \cite{Kume05}, holonomic gradient descent \cite{Koyama2012}, and precomputed lookup tables \cite{Glover13}.  The latter approach is used in this work, because it offers a fast evaluation and thus, makes an efficient implementation of the filter possible. This approach is also chosen for computing derivatives of the normalization constant, which will be needed in parameter estimation.

 The product of two Bingham pdfs is again a (rescaled) Bingham pdf. The proof is similar to the Gaussian case and presented in \cite{Fusion13_Kurz-Bingham}. Furthermore, for a $\binghamd(\fM,\fZ)$-distribution, the identity matrix $\fI$ and all $c\in\R$ we have
\begin{align}\label{eq:transformZMatrix}
\binghamd(\fM, \fZ) = \binghamd(\fM, \fZ+ c\fI)\ .
\end{align}
Even though a Bingham distributed random vector $\uy$ only takes values on the unit sphere, it is still possible to compute a classical (i.e., non-circular) covariance matrix in $\R^d$, which is given by 
\begin{align*} 
\Cov(\uy) =\, \E(\uy^2) - \underbrace{(\E(\uy))^2}_{0} 
 =\,  \fM \cdot \diag\li(
\fr{\fr{\partial }{\partial z_1}N(\fZ)}{N(\fZ)}, \ldots,
\fr{\fr{\partial }{\partial z_d}N(\fZ)}{N(\fZ)}
\ri) \cdot \fM^T
\end{align*} 
according to \cite{Bingham1974}. Following the definition of the Bingham distribution, this is simply the covariance matrix of a normal distributed random vector $\ux\sim\normd_d(\uzero, -0.5(\fM(\fZ+c\fI)\fM^T)^{-1})$ given $\norm{\ux}=1$, where $c\in\R$ can be chosen arbitrarily as long as  $(\fM(\fZ+c\fI)\fM^T)$ is negative definite.

For estimating parameters of a Bingham distribution based on moment matching, we consider a given covariance matrix $\fC$ from a spherical distribution with mean $\uzero$. Let ${\fM\cdot\diag(\om_1, \ldots\om_d)\cdot\fM^T}$ be the eigendecomposition of $\fC$, where the columns of $\fM$  consist of orthogonal eigenvectors. 
Solving 
\begin{align}\label{eq:binghamAxes} 
\fr{\fr{\partial }{\partial z_i}N(\fZ)}{N(\fZ)} = \om_i\ ,
\qquad i=1,\ldots,d
\end{align}
yields $\fZ = \diag(z_1, \ldots, z_d)$. From \eqref{eq:transformZMatrix}, it can be seen that one of the parameters $z_d$ can be chosen arbitrarily (for computational reasons, the matrix $\fZ$ is usually chosen so that $z_d=0$). The resulting $\binghamd(\fM,\fZ)$ distribution has covariance $\fC$.

\begin{figure}
\begin{center}
\includegraphics[width=0.6\columnwidth]{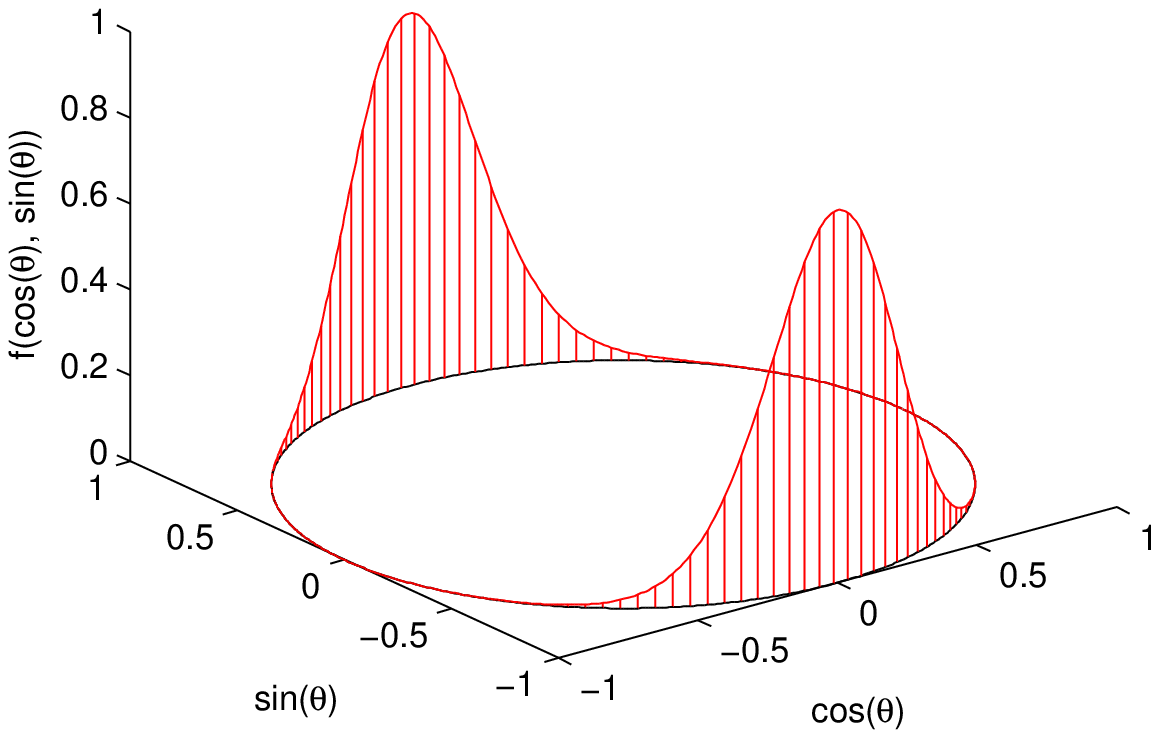}
\caption{The pdf of a two-dimensional Bingham distribution, which is defined on the circle.} 
\end{center}
\end{figure}

Typically, orientations can be represented by unit quaternions \cite{Kuipers2002}. A clockwise rotation of $\te$ degrees around the unit length axis $\ua = (a_x, a_y, a_z)^T$ is represented by the quaternion   
\begin{align*}
\fq = \cos\li(\fr{\te}{2}\ri) 
  + (a_x i + a_y j + a_z k)\sin\li(\fr{\te}{2}\ri)\ .
\end{align*}
The orientations represented by the quaternions $q$ and $-q$ are identical. Thus, a Bingham distribution on $S^3$ is suitable for representing uncertainty  of the quaternion ${\fq=q_1 + q_2 i + q_3 j + q_4k}$, which will be represented by the Bingham distributed random vector $\uy=(q_2, q_3, q_4, q_1)^T$. Putting the real part of the quaternion $a_1$ at the last position of our vector is due to the fact, that the Bingham distribution has its maxima at $\pm \fM^T (0,0,0,1)^T$.

The composition of rotations can be represented by a quaternion multiplication (Hamilton product). In the case of our Bingham random vectors, we denote this operation by $\oplus$, which yields for two Bingham random vectors $\ux$ and $\uy$
\begin{align*}
\bpmat x_1\\ x_2\\ x_3\\ x_4 \epmat \oplus
\bpmat y_1\\ y_2\\ y_3\\ y_4 \epmat := \bpmat
x_4y_1 + x_1y_4 + x_2y_3 - x_3y_2\\
x_4y_2 - x_1y_3 + x_2y_4 + x_3y_1\\
x_4y_3 + x_1y_2 - x_2y_1 + x_3y_4\\
x_4y_4 - x_1y_1 - x_2y_2 - x_3y_3
\epmat\ .
\end{align*}
Unfortunately, the resulting random vector does not follow a Bingham distribution. Moment matching is used to approximate $\ux\oplus\uy$ by a Bingham distribution. $\Cov(\ux\oplus\uy)$ can be computed directly as in \cite{Fusion13_Kurz-Bingham}, \cite{Glover2013} and the approximation step is performed by computing an eigendecomposition of this covariance matrix and then solving eq.~\eqref{eq:binghamAxes}. This yields the parameters of our approximated Bingham distribution.

\begin{figure*}
\begin{center}
\begin{subfigure}{0.32\textwidth}
\centering
\includegraphics[width=\textwidth]{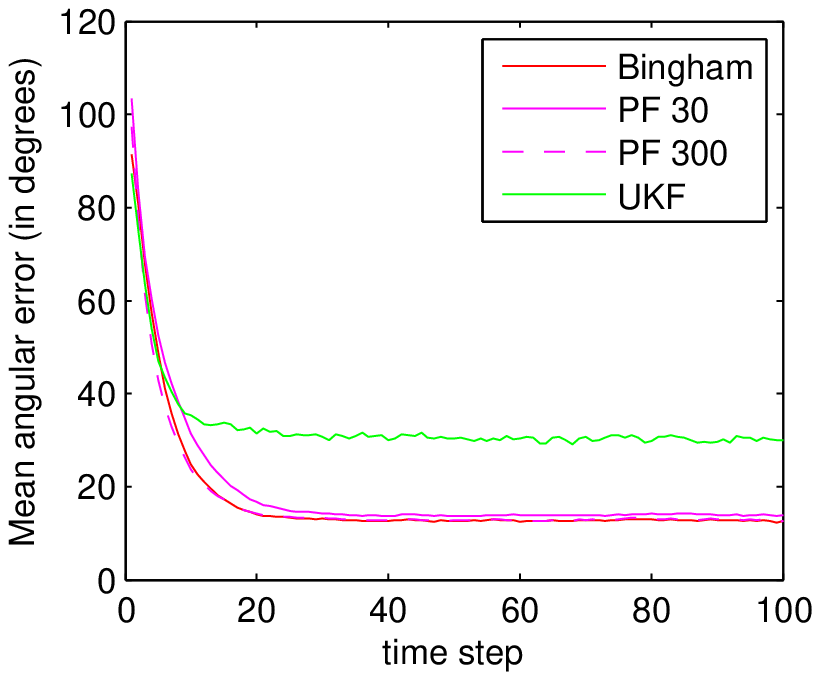}
\caption{Mean angular error. \\$\quad$}
\end{subfigure}\hfill
\begin{subfigure}{0.32\textwidth}
\centering
\includegraphics[width=\textwidth]{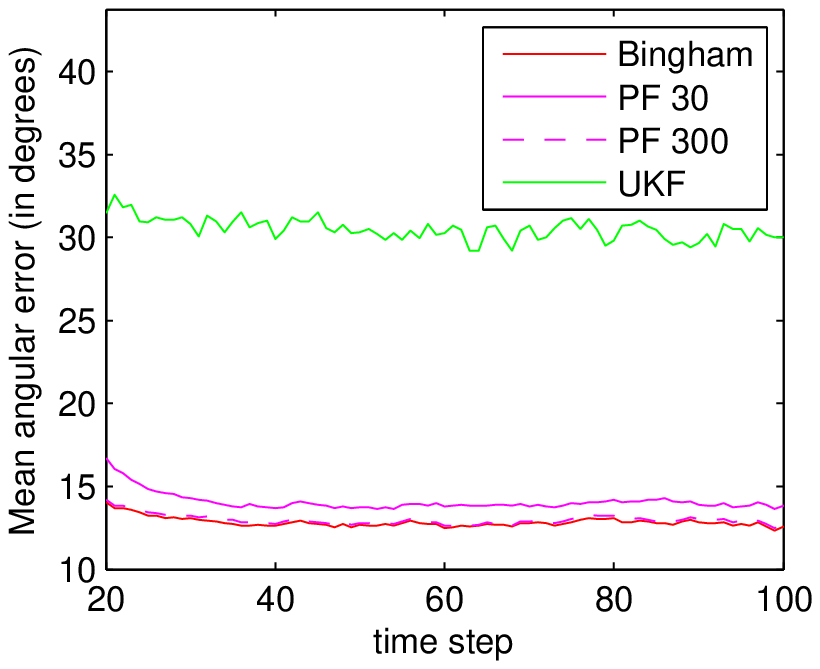}
\caption{Mean angular error after initialization.}
\end{subfigure}\hfill
\begin{subfigure}{0.32\textwidth}
\centering
\includegraphics[width=\textwidth]{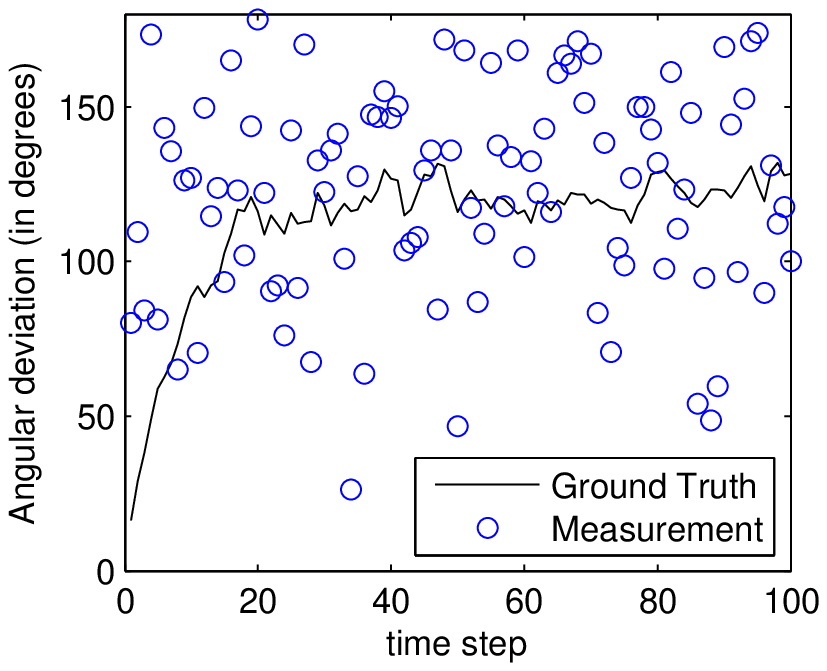}
\caption{A typical run.\\$\quad$}
\end{subfigure}
\caption{\label{fig:eval}Error evaluation after 1000 Monte Carlo runs in the case involving high measurement noise. The typical run is given as the angle between the true system state and the quaternion $(0,0,0,1)$, respectively $(0,0,0,-1)$. The error is given as the angle between the true system state and the estimates of the considered filters.}
\end{center} 
\end{figure*}

\section{Deterministic Sampling of Uncertain Orientation}\label{sec:sampling}\noindent
Our use of the Bingham distribution is motivated by its ability to represent uncertainty over orientations  parametrized by unit quaternions. Only simple transformations of such uncertain quaternions preserve the Bingham distribution, for example change of the current orientation in a predefined direction without introducing any further noise.  

Computing the transformation $g(\uy)$ of a $\binghamd(\fM, \fZ)$ distributed random variable $\uy$ is not possible in closed form for arbitrary functions $g(\cdot)$. Thus, we propose a technique to approximate a Bingham distribution on $S^{d-1}$ by $4d-2$ deterministically placed samples adapting the basic idea of the Unscented Transform to the manifold of orientations. With regard to quaternions, we consider the case $d=4$. Thus, each sample can be considered as a quaternion describing an orientation. One sample is placed at the pole $(0,0,0,1)^T$, which can be thought of as a mode on the sphere. Furthermore, six samples  
\begin{align*}
\bpmat \pm \sin(\al_1) \\ 0 \\ 0 \\ \cos(\al_1) \epmat\ ,\ 
\bpmat 0 \\ \pm \sin( \al_2) \\ 0 \\ \cos(\al_2) \epmat\ ,\ 
\bpmat 0 \\ 0 \\ \pm \sin( \al_3) \\ \cos(\al_3) \epmat\ 
\end{align*} 
are placed around this pole. Negation yields samples around the pole $(0,0,0,-1)$ to account for antipodal symmetry. That is, we obtain another seven samples which are mirror images of the first set. Each pole is assigned the probability mass $p_0/2$ and each sample corresponding to the angle $\al_i$ is assigned the probability mass $p_i/4$. Each sample is multiplied by $\fM$. Thus, the sample based probability distribution generated by this method has covariance $\fM\fC\fM^T$ where $\fC$ is given by
\begin{align*} 
\fC=\diag\Bigg( p_1\sin(\al_1)^2,\ p_2\sin(\al_2)^2,\ 
p_3\sin(\al_3)^2,\ 
p_0 + \sum_{i=1}^3 p_i\cos(\al_i)^2 \Bigg)\ . 
\end{align*} 
In the next step, we use moment matching to find $\al_i$ and $p_i$ so that our samples have the same uncertainty as the approximated Bingham distribution. This is done by solving
\begin{align}\label{eq:matchMoments}
\fC=\diag(\om_1, \ldots,\om_4)\ ,
\end{align} 
with $\om_i$ as defined in \eqref{eq:binghamAxes}, which gives 
\begin{align*}
\al_i = \arcsin\li(\sqrt{\fr{\om_i}{p_i}}\ri)\ ,\qquad i=1,2,3\ .
\end{align*}
Thus, we require $\om_i\leq p_i$. Choosing
\begin{align*}
p_0 =\ & \la\om_4\ ,\\
p_i =\ & \om_i + (1-\la) \fr{\om_4}{3}\ ,\qquad i=1,2,3
\end{align*} 
gives a feasible solution to \eqref{eq:matchMoments} for every $\la\in [0,1)$. Finally, our deterministically sampled distribution has covariance ${\fM\cdot\diag(\om_1, \ldots, \om_4)\cdot\fM^T}$, which corresponds to the covariance of our Bingham distribution and completes the approximation. 

\section{Quaternion Based Filtering of Orientation}\label{sec:filter}\noindent
Our recursive filter is separated into a prediction step, and a filter step. Currently, the measurement update step considers noisy but direct measurements of the true system state. The prediction step uses deterministic sampling. Thus, it can be seen as a counterpart to the UKF.


\subsection{Prediction Step}
We consider functions $g: S^3 \mapsto S^3$ with $g(\ux)=-g(-\ux)$ to ensure the antipodal symmetry. After propagation through the system function some orientational noise is imposed to account for the uncertainty. Thus, the considered system function is given by
\begin{align*}
\ux_{t+1} = g(\ux_t) \oplus \uw_t\ ,
\end{align*}
where $\uw_t \sim \binghamd(\fM^w_t, \fZ^w_t)$. We use deterministic sampling as described in the preceding section to approximate our current system state $\ux_t \sim\binghamd(\fM_t^e, \fZ_t^e)$. 
\begin{enumerate}
  \item Approximate $\binghamd(\fM_t^e, \fZ_t^e)$ using deterministic sampling.
  \item Propagate each sample through the system function $g(\cdot)$.
  \item Compute approximation of $\Cov(g(\ux_t))$ by computing the sample covariance after propagation.
  \item Use this approximated covariance to compute $\Cov(g(\ux_t)\oplus\uw_t)$ and obtain $\fM^p_{t+1}$ and $\fZ^p_{t+1}$ by moment matching. 
\end{enumerate}
For more general system models, such as $\ux_{t+1}=g(\ux_t, \uw_t)$ one would replace steps 3) and 4) by deterministic sampling of $\ux_t$ and $\uw_t$, and approximating $\Cov(g(\ux_t, \uw_t))$ by the sample covariance. After this procedure our predicted system state is described by a $\binghamd(\fM^p_{t+1},\fZ^p_{t+1})$ distribution and the corresponding maximum likelihood estimate $\pm\hux_{t+1}^p$ of the orientation is described by the last column of $\pm\fM^p_{t+1}$ (because the order of the entries in $\fZ^p_{t+1}$ was chosen in an increasing order and $z_d=0$).

\subsection{Measurement Update Step}
Our measurement equation is given by 
\begin{align*}
\uz_t = \ux_t \oplus \uv_t\ ,
\end{align*}
where $\uv_t\sim\binghamd(\fM^v_t, \fZ^v_t)$. Choosing the identity matrix as the first parameter of the Bingham distribution, i.e. $\fM^v_t = \fI$, corresponds to the concept of zero-mean noise in the Euclidean space. This results in the Bayesian estimator
\begin{align*}\label{eq:productOfBinghams}
f(\ux_t|\huz_t) 
  = c \cdot f(\huz_t | \ux_t) \cdot f(\ux_t)\ ,
\end{align*} 
where $c$ is a normalization constant. Here, $f(\ux_t)$ is the pdf of a $\binghamd(\fM_t^p, \fZ_t^p)$ distribution and
\begin{align*}
f(\huz_t | \ux_t) 
  = f(\ux^{-1}_t \oplus \huz_t; \fM^v_t, \fZ^v_t) 
  = f(\ux_t ; \bfM_t^v \oplus \huz_t, \fZ^v_t )
\end{align*}
is the pdf of a $\binghamd(\bfM_t^v \oplus \huz_t, \fZ^v_t)$ distribution, where $\ux^{-1}_t$ denotes the inverse rotation to the rotation represented by $\ux$, $\bfM$ denotes the matrix $\fM$ where the first three rows are multiplied by $-1$ (to account for a conjugated quaternion) and $\bfM_t^p \oplus \huz_t$ denotes a matrix resulting from applying the Hamilton product $\oplus$ to each column vector. As already mentioned in the preceding section, we make use of the fact that the product of two Bingham pdfs is again a rescaled Bingham pdf and thus the measurement update yields 
\begin{align*} 
f(\ux_t|\huz_t) \propto  
 \quad \exp(\ux_t^T (\underbrace{ 
  (\bfM_t^v \oplus \huz_t) \fZ^v_t (\bfM_t^v \oplus \huz_t)^T
   + \fM^p_t \fZ_t^p (\fM^p_t)^T}_{\fA:=} )\ux_t)\ .
\end{align*}
Applying an eigendecomposition to $\fA$ yields the parameters ($\fM^e_t$, $\fZ^e_t$) describing our posterior Bingham distribution. Thus, the whole measurement update is performed by computing the matrix $\fA$ and its eigendecomposition.

\section{Evaluation}\label{sec:evaluation}\noindent
\begin{figure*}[t]
\begin{center}
\begin{subfigure}{0.4\textwidth}
\centering
\includegraphics[width=0.8\columnwidth]{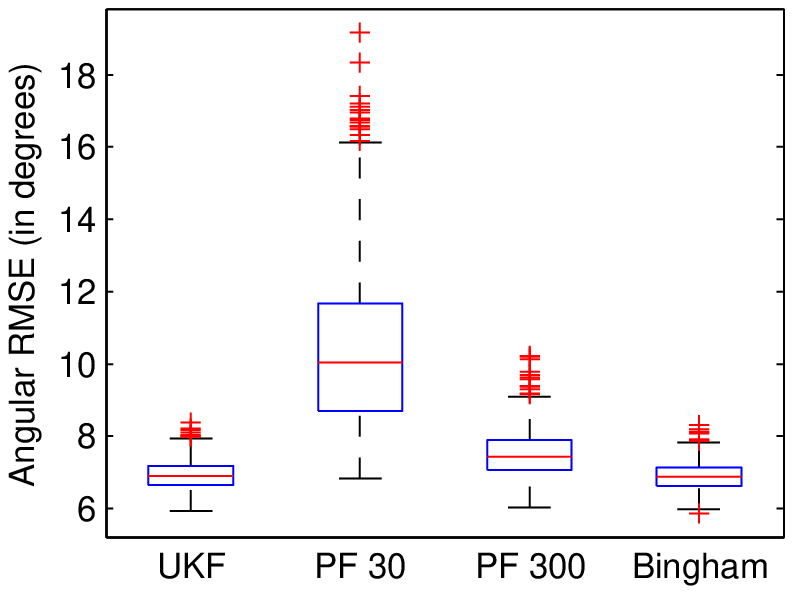}
\caption{Low noise.}
\end{subfigure}
\begin{subfigure}{0.4\textwidth}
\centering
\includegraphics[width=0.8\columnwidth]{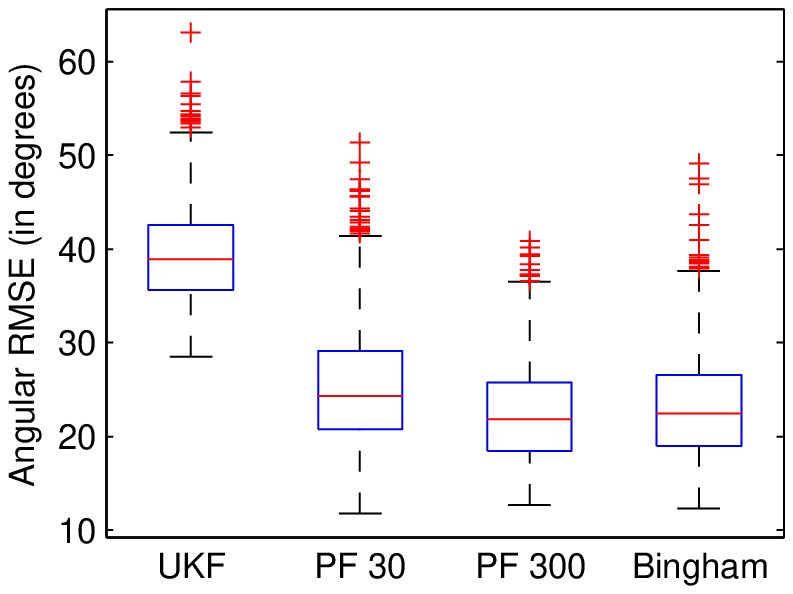}
\caption{High noise.}
\end{subfigure}
\end{center}
\caption{\label{fig:results}The proposed filter performs better for strong noise by taking its periodic nature into account. The UKF uses 9 samples, while the particle filter uses up to 100 in our simulation. By considering antipodal symmetry, it is sufficient for the proposed filter to use 7 samples.}
\end{figure*}
We consider a stabilization scenario of a robotic ball joint in order to evaluate the proposed filtering approach. This problem arises in servoing a sensor on a moving platform to always point in a given direction. Thus, in this scenario, the true system state is stabilized towards a predefined goal state $\uy$. The magnitude of stabilization feedback imposed by the system depends on the actual deviation from this goal state. Furthermore, a measurement model involving noisy measurements is considered. This results in system and measurement models given by  
\begin{align*}
g(\ux_t,\uw_t) =\ & \ux_t \oplus (\ux_t^{-1}\oplus\uy)^u\oplus \fr{\uw_t}{\norm{\uw_t}}\ ,\\
h(\ux_t, \uv_t) =\ & \ux_t \oplus \fr{\uv_t}{\norm{\uv_t}}\ ,
\end{align*}
where $\uw_t \sim \normd(\umu_w, \fC_w)$, $\uv_t \sim \normd(\umu_v, \fC_v)$, and the parameter $u$ controls the magnitude of the system model. Here, the use of Gaussian noise is motivated by avoiding an unfair advantage of the Bingham filter. Computation of $\ux_t^{-1}$ and $(\ux_t^{-1}\uy)^u$ is performed by interpreting the vectors as quaternions. The considered system function also appears in the context of spherical interpolation \cite{Shoemake1985}.

The proposed filter was compared against a modified UKF and a modified particle filter with importance resampling after each update step. In order to further help the UKF and the particle filter to handle antipodal symmetry, the measurement is checked and if necessary multiplied by $-1$ to ensure $\norm{\uz_t-\ux^p_t }<\norm{-\uz_t-\ux^p_t}$. For the particle filter, this check needs to be done for each particle. In the case of the UKF, it is sufficient to perform this check once in each measurement update step. Furthermore, the measurement likelihood in the particle filter is approximated by a Gaussian distribution according to 
\begin{align*}
p(\uz_t|\ux_t) \approx f(\ux_t^{-1}\oplus\huz_t)\ ,
\end{align*}
where $f(\cdot)$ denotes the pdf of $\uv_t$ and $\ux_t^{-1}$ is the vector representing the inverse quaternion of $\ux_t$. Two instances of the particle filter (with 30 and 300 particles) were used.

The proposed filter assumes $\uw_t/\norm{\uw_t}$ and $\uv_t/\norm{\uv_t}$ to be Bingham distributed. Thus, the parameters  $(\fZ^w, \fM^w)$ and $(\fZ^v, \fM^v)$ need to be matched to  our system model. This can be done by generating random samples for each Gaussian distribution involved (we used 10\,000 samples). Afterwards, these samples are normalized to length 1 and the Bingham distribution parameters are matched as described in Sec. II. 

Using the classical RMSE as an error measure would be misleading in several ways. First, it does not sufficiently consider the spherical nature of our probability distribution. Second, it does not consider antipodal symmetry and would consider $q$ to be a wrong estimate of $-q$ even though both represent the same rotation. Both problems are tackled by introducing an angular error
\begin{align*}
\al_i = 2\cdot \min(\acos\li((\ux_t)^T\cdot \ux_t^e \ri),
\pi-\acos\li((\ux_t)^T\cdot \ux_t^e \ri))\ .
\end{align*}
Here, the minimization procedure accounts for antipodal symmetry. This definition is used for computing the mean angular error and the angular RMSE. The angular error corresponds to the angle in the Rodrigues rotation formula for a rotation from $\ux_t$ to $\ux_t^e$, which can be seen by considering two arbitrary quaternions $\fa, \fb\in\Hbb$. We know, that the the Rodrigues angle describing a transformation from orientation $\fa$ into orientation $\fb$ is given by $\te:=2\cdot\cos\li(\Re\li(\fb\oplus \fa^{-1}\ri)\ri)$, where $\oplus$ once again denotes the classical quaternion multiplication, i.e., the Hamilton product. Using $\ua$ and $\ub$ as vector representations for the quaternions $\fa$, $\fb$, it follows
\begin{align*}
\te / 2 
 =\ & \acos\li(\Re\li(\fb\oplus\fa^{-1}\ri) \ri)\\
 =\ & \acos( b_4a_4-b_1(-a_1) - b_2(-a_2)-b_3(-a_3) ) \\
 =\ &\acos(\ub^T\ua)\ .
\end{align*}
The proposed angular notion is also used to plot the true system state evolution as the angular deviation from $(0,0,0,1)^T$. This vector is chosen, because it represents the identity in the skew-field of quaternions and thus stands for no change of orientation.

The initial parameters used to generate the ground truth were $\umu_0=\umu_w=\umu_v=(0,0,0,1)^T$, $\fC_0 = 0.01$, and $\fC_w = 0.001\cdot\fI$ (which corresponds to an expected angular deviation of $18.2^\circ$ and $5.8^\circ$ respectively). The covariance of measurement noise was $\fC_v=0.003\cdot\fI$ for a low noise scenario and $\fC_v=0.3\cdot\fI$ for a high noise scenario (corresponding to an expected angular deviation of $10.0^\circ$ and $85.8^\circ$ respectively). Here, $\fI$ once again denotes the identity matrix. The goal region was chosen as $\uy=(0.5, 0.5, 0.5, 0.5)^T$ and the exponent $u$ was chosen as $0.1$. An initial estimate was given by $\umu_e=(1, 0, 0, 0)^T$ and $\fC_e=\fI$. The corresponding Bingham distribution parameters for the initial estimate were found in the same way as described above. For deterministic sampling in the prediction step of the proposed filter, we used $\la=0.5$.  

Using this setup, 1\,000 Monte Carlo runs were performed for both measurement noise scenarios described above. Each run simulated 100 time steps. The mean angular error in each time step and ground truth of a typical run (in the high noise scenario) are shown in Fig.~\ref{fig:eval}. Evaluations of the angular RMSE are shown in Fig.~\ref{fig:results}. The proposed approach outperforms the UKF and the particle filters. This effect becomes smaller for high certainty and stronger for a higher noise in the measurement model. 

We compared the computation time using Matlab 2013a on a system using an Intel i7-2620M processor. The better estimation results of the Bingham based filter come at the price of higher computation time. In our setup, the average computation times were 305ms for one Bingham filter step, 317ms for one  particle filter step (with 300 particles), and 27ms for the UKF. The libbingham \cite{Glover13} statistics library was used for computing the Bingham normalization constant and its derivatives. Higher computation time of the Bingham filter is due to a numerical moment matching procedure.


\section{Conclusions}\label{sec:conclusion}\noindent
Earlier approaches to tackle the problem of developing recursive filters for dynamic estimation of angular and orientational data focused on coming up with better representations of the system state or a better consideration of the system function. The use of directional statistics makes a better consideration of the stochastic uncertainties in this kind of estimation problems possible. Particularly applications involving strong system and measurement noise will benefit from this development.

In this work, a recursive filter for orientation estimation based on the Bingham distribution was proposed. It involves deterministic sampling and can be considered as a hypersphere based equivalent to the UKF. Particularly, for scenarios with strong noise, the proposed approach outperforms classical filtering techniques based on the Gaussian distribution. Our future work might involve more efficient computational techniques, measurement update algorithms for more complex measurement functions, and using deterministic sampling for a combination of directional and non-directional quantities including platform position or inertial sensor measurement biases.


\section*{Acknowledgment}
This work was partially supported by grants from the German Research Foundation (DFG) within the Research Training Groups RTG 1194 ``Self-organizing Sensor-Actuator-Networks'', and RTG 1126 ``Soft-tissue Surgery: New Computer-based Methods for the Future Workplace'', as well as the European Commission within the BEAMING project.


\newpage

\bibliographystyle{IEEEtranNoUrl}
\bibliography{ig-letter,gk-letter,ISASPublikationen_laufend}

\end{document}